\documentclass[twocolumn,showpacs,preprintnumbers,amsmath,amssymb]{revtex4}
\usepackage{graphicx}
\usepackage{dcolumn}
\usepackage{bm}
\newcommand{\be}{\begin{equation}}
\newcommand{\ee}{\end{equation}}
\newcommand{\ba}{\begin{eqnarray}}
\newcommand{\ea}{\end{eqnarray}}

\begin{document}
\preprint{IC/2006/43}
\preprint{MCTP-06-12}
\title{$M$ theory Solution to the Hierarchy Problem}
\author{Bobby Acharya}
\affiliation{%
Abdus Salam ICTP, Strada Costiera 11, Trieste, Italy}%
\author{Konstantin Bobkov}%
\author{Gordon Kane}
\author{Piyush Kumar}
\author{Diana Vaman}
\affiliation{MCTP, University of Michigan, Ann Arbor, MI, USA}%
\date{\today}
\begin{abstract}
An old idea for explaining the hierarchy is strong gauge dynamics.
We show that such dynamics {\it also} stabilizes the moduli in $M$ theory
compactifications on manifolds of $G_2$-holonomy {\it without} fluxes. This gives
stable vacua with softly broken susy, grand unification and a distinctive spectrum of TeV
and sub-TeV sparticle masses.
\end{abstract}
\maketitle
\noindent{\bf 1. Stabilizing Hierarchies and Moduli}\\
$M$ theory (and its weakly coupled string limits) is a consistent quantum
theory including gravity, particle physics and much more. Although apparently unique,
the theory has a large number of solutions, manifested by the presence of moduli:
massless scalar fields with classically undetermined
vacuum expectation values (vevs), whose values determine the masses and coupling constants of the
low energy physics.

In recent years, there has been substantial progress in understanding mechanisms which stabilize
moduli in various corners of the $M$ theory moduli space. In particular, the stabilization of all
moduli by magnetic fields (fluxes) in the extra dimensions, perhaps also combined
with other quantum effects, has been reasonably well understood in the context of Type IIB string theory
\cite{Giddings:2001yu,Kachru:2003aw}, $M$ theory \cite{Acharya:2002kv} and Type IIA string theory \cite{DeWolfe:2005uu}.
After stabilizing the moduli, one still has to explain why
$M_{W}/m_{pl} \sim 10^{-16}$.

The effective potential of these compactifications fits into the framework of a low energy
supergravity theory in four dimensions. A well known property of the latter is that there is a universal
contribution to scalar masses of order the gravitino mass $m_{3/2}$. Therefore, without miraculous cancellations,
in theories in which $m_{3/2}$ is large, the Higgs mass is also large. In $M$ theory and Type IIA
flux vacua the vacuum superpotential is $\mathcal{O}(1)$ or larger in Planck units. This gives a large $m_{3/2}$
(unless the volume of the extra dimensions is large, ruining standard unification). In heterotic flux vacua
\cite{Gukov:2003cy} $m_{3/2}$ can be smaller, but only by a few orders of magnitude.
Thus, in these vacua, stabilizing
the moduli using fluxes {\it fails to solve the hierarchy problem}, {\it viz.} to generate and stabilize the hierarchy
between the electroweak and Planck scales.

In Type IIB theory, this is not so: $m_{3/2}$ can be tuned small by choosing fluxes.
One can also address the possibility
of generating the hierarchy through warping \cite{Randall:1999ee} in this framework \cite{Giddings:2001yu}.
The hierarchy problem is less well understood in other corners of the $M$ theory moduli space.

Our focus will be $M$ theory, and we will henceforth
switch off all the
fluxes else the hierarchy will be destroyed.
Supersymmetry then implies that the seven extra dimensions form a space $X$ with $G_2$-holonomy.
In these vacua,
non-Abelian gauge fields are {\it localized} along three dimensional submanifolds $Q$ $\subset X$ at which there is an orbifold
singularity \cite{Acharya:1998pm}
and chiral fermions are {\it localized} at points at which there are conical singularities \cite{Atiyah:2001qf,Acharya:2004qe,Acharya:2001gy}.

These vacua can have interesting phenomenological features, independently of how moduli are stabilized:
the Yukawa couplings are hierarchical; proton decay proceeds at dimension six with distinctive decays;
grand unification is very natural; the
$\mu$-term is zero in the high scale lagrangian \cite{Witten:2001bf, Acharya:2001gy,Friedmann:2002ty, Acharya:2005ks}.
Also, since the $Q$'s generically do not intersect each other, {\it supersymmetry breaking will be gravity mediated in these vacua}.
Therefore, it is of considerable interest to understand whether or not there exist mechanisms
which can a) stabilize the moduli of such compactifications, b) generate a hierarchy of scales, and if so, c)
what is the resulting structure of the
soft terms and their implications for LHC?

All the moduli fields $s_i$ have axionic superpartners $t_i$, which, in the absence of fluxes, enjoy a Peccei-Quinn shift symmetry.
This is an important difference with respect to other $M$ theory limits such as heterotic or Type IIB.
Therefore, in the zero flux sector, the {\it only contributions to the superpotential are non-perturbative}. These can arise either
from strong gauge dynamics or from membrane instantons. Since the theory of membrane instantons in $G_2$ manifolds is technically challenging \cite{Harvey:1999as}, we will restrict
our attention to the strong gauge dynamics case henceforth.

Furthermore, unlike its weakly coupled string limits, in $M$ theory the
non-perturbative superpotential in general {\it depends upon all the moduli}.
Hence, one would expect that the effective supergravity potential has isolated minima.
Our main conclusion is that strong gauge dynamics produces an effective potential which indeed {\it stabilizes all moduli
and generates an exponential hierarchy of scales}. After describing this result, we also briefly describe the pattern of
soft breaking terms which these vacua predict and begin to discuss the consequences for the LHC.

\noindent{\bf 2. The Moduli Potential} \\
The moduli Kahler potential is difficult to calculate explicitly.
However, a family of Kahler potentials, consistent with
$G_2$-holonomy and known to describe accurately some explicit
examples of $G_2$ moduli dynamics were given in
\cite{Acharya:2005ez}.  These are defined by \be \label{kahler} K
= -3 \ln(4\pi^{1/3}\,V_X),\,V_X = \prod_{i=1}^{N} s_i^{a_i},\,
\mathrm{with}\,\sum_{i=1}^{N} a_i = 7/3.\ee where $V_X$ is the
volume of the $G_2$ holonomy manifold as a function of the $N$
scalar moduli $s_i$(in 11d units). The superpotential for the
simple case of a hidden sector without charged matter is:

Therefore,
\begin{eqnarray}
\label{generalsuper}
W = \sum_{k=1}^{M}A_ke^{ib_k f_{k}},\,\; f_k=\sum_{i=1}^{N} N^k_i\,z_i\,=
\frac{\theta_k}{2\pi}+i\frac{4\pi}{g_k2}\,.
\end{eqnarray}
\noindent $M$ is the number of hidden sectors whose gauginos condense,
$b_k=\frac {2\pi}{c_k}$ with $c_k$ the
dual coxeter number of the $k$-th gauge group whose 4d gauge coupling function
$f_k$ is an integer linear combination of the moduli fields $z_i=t_i+i s_i$.
The $A_k$ are (RG-scheme dependent) numerical constants. More general cases will be described in \cite{Toappear}.

Note that all of the `parameters' which enter the potential, i.e.
$( b_k , A_k , N^k_i )$, are constants. $b_k$ and $N^k_i$ are
straightforward to determine from the topology of $X$. The one
loop factor $A_k$ is more difficult to obtain, but e.g. the
threshold corrections calculated in \cite{Friedmann:2002ty} show
that they can be computed and can take a reasonably wide range of
values in $M$ theory.

At this point the simplest possibility would be to consider a single hidden sector gauge group.
While this {\it does in fact stabilize all the moduli}, it is a) non-generic and b) fixes
the moduli in a place which is strictly beyond the supergravity approximation.
Therefore we will consider two such hidden sectors, which is more representative of
a typical $G_2$ compactification as well as being tractable enough to analyze.
The superpotential therefore has the following form
\be
\label{super} W^{np} = A_1e^{ib_1 f_1}+ A_2e^{ib_2 f_2}\,. \ee

The scalar potential can be computed from $K$ and $W$, and after integrating
out the axions (without loss of generality we chose $A_k > 0$),
it is given by, in 4d Planck units,
\begin{widetext}
\ba \label{potential}
V=\frac{1}{48\pi V_X3}\,[\sum_{k=1}^{2}\sum_{i=1}^{N}a_i{\nu_i^k}\left({\nu_i^k}b_k+3\right)b_kA_k^2e^{-2b_k\vec\nu^{\,k}\cdot\,\vec a}+3\sum_{k=1}^{2}A_k^2e^{-2b_k\vec\nu^{\,k}\cdot\,\vec a}&&\\
-2\sum_{i=1}^{N}a_i\prod_{k=1}2\nu_i^kb_kA_ke^{-b_k\vec\nu^{\,k}\cdot\,\vec a}
-3\left(2+\sum_{k=1}^{2}b_k\,\vec\nu^{\,k}\cdot\vec a \right)\prod_{j=1}^2A_je^{-b_j\vec\nu^{\,j}\cdot\,\vec a}]&&\,,\nonumber
\ea
\end{widetext}
where we introduced a variable
\be \label{nu} \nu_i^k\equiv\frac{N_i^k s_i}{a_i} \;\;(\mathrm{no\;\,sum});\;\;{\rm Im}f_k=\vec\nu^{\,k}\cdot\,\vec a\,.
\ee
\noindent{\bf 2.1 Vacua} \\
Vacua of the theory correspond to stable critical points of the potential.
Although, as we will see, the potential has stable vacua with spontaneously broken supersymmetry,
it is instructive to analyze the supersymmetric vacua.
For simplicity we will describe here only
the special case when the two groups have the same gauge coupling (explicit examples are given in sect. 3).
See \cite{Toappear} for the more elaborate general case.

In this special case, we have
\be \label{ratio} N^1_i=N^2_i=N_i \implies \nu^1_i=\nu^2_i=\nu_i\equiv\frac{N_i
s_i}{a_i}.\, \ee As a result, the $F$-terms
($F_i=\partial_iW+(\partial_iK)W$) simplify
significantly. Solving  $F_i=0$ yields: \ba\label{eq8} \nu_i \equiv \nu =
-\frac{3(\alpha-1)}{2(\alpha b_1-b_2)},\,\mathrm{with}\,\frac{A_2}{A_1}\, =\frac 1 \alpha
e^{\frac 7 2(b_1-b_2)\frac{(\alpha-1)}{(\alpha b_1-b_2)}}\ea
where $\alpha$ is determined by the right equation. Since $\nu_i$ is independent of $i$, it is also independent of
the number of moduli $N$, \emph{which means that this solution
fixes all moduli for a manifold with any number of moduli}.

In the limit of large $\nu$, $\nu$ is given by (for gauge groups
$SU(P)$ and $SU(Q)$) \be \label{log} \nu \sim {3 \over 7(b_2 -
b_1)}\log{A_2 b_2 \over A_1 b_1} = {3 \over 14\pi}{PQ \over
P-Q}\log{A_2 P \over A_1 Q} \ee

This is a very good approximation for $\nu >
O(1)$, and shows that the moduli vevs can be greater than one for
gauge group ranks less than 10, yielding solutions within the
supergravity approximation. However, there will be an upper bound
on the moduli vevs in these vacua, since we expect that $A_1, A_2,
P, Q$ have upper limits. The dependence of (\ref{log}) on the
input parameters is similar to those obtained for other
constructions \cite{deCarlos:1992da}. Once $\nu$ is determined,
the moduli $s_i$ are found from (\ref{ratio}) and the hierarchy
between the moduli vevs is determined by the ratios $a_i/N_i$.
For cases when $PA_2/QA_1$ is order one,
it is not clear if additional corrections
change the results significantly. Similar issues were faced in \cite{deCarlos:1992da,ddf}.

\noindent{\bf 2.2 Minima with spontaneously broken supersymmetry}
Formally, the potential has $2^{N} -1$ extrema with spontaneously broken susy
and one supersymmetric one \cite{Toappear}.
For simplicity we will exhibit these for the two moduli case.
For example, consider the parameter set\begin{eqnarray}
\label{choice1}\{A_1,A_2,b_1,b_2,N_1,N_2,a_1,a_2\}=\{0.12,2,\frac{2\pi}{8},\frac{2\pi}{7},1,1,\frac{7}{6},\frac{7}{6}\}
\nonumber
\end{eqnarray}
\noindent The solutions are :
\begin{eqnarray} \label{sol7} s^{(1)}_1&\approx&13.05\,,\,\,\,s^{(1)}_2\approx13.05\,\;\mathrm{\footnotesize{(susy\;extremum)}} \\
s^{(2)}_2&\approx&13.59\,,\,\,\,s^{(2)}_2\approx13.59\,\;\mathrm{(de\,Sitter\;extremum)}\nonumber\\
s^{(3)}_1&\approx&2.61\,,\,\,s^{(3)}_2\approx23.55\,\;\mathrm{(non\,susy\;AdS\;minimum)}\nonumber\\
s^{(4)}_1&\approx&23.55\,,\,\,s^{(4)}_2\approx2.61\,\; \mathrm{(non\,susy\;AdS\;minimum)}\nonumber\,.
\end{eqnarray}



The supersymmetric extremum in
(\ref{sol7}) is a saddle
point. The two stable minima spontaneously break
supersymmetry. Visual plots of the potential can be seen at \cite{web}. The stable minima appear
symmetrically though generically, for $a_1\not=a_2$ and/or
$N_1\not =N_2$ one of the minima will be
deeper than the other. For the case under investigation,
the volume is stabilized at the value
$V_X=122.3$ which is presumably large enough for the supergravity analysis to hold.

\noindent{\bf 3. Explicit Examples}\\
To prove the existence of a $G_2$-holonomy metric on a compact 7-manifold $X$ is a difficult
problem.
There is no analogue of Yau's theorem for Calabi-Yau manifolds
which allows an ``algebraic'' construction.
Nevertheless, Joyce and Kovalev have successfully constructed many smooth examples \cite{Kovalev:2001zr}.
Furthermore, dualities with heterotic and Type IIA string vacua also imply the existence of many
singular examples. The vacua discussed here have {\it two} gauge groups
so $X$ will have {\it two} submanifolds $Q_1$ and $Q_2$ of orbifold singularities.

Kovalev constructs $G_2$ manifolds which can be described as the
total space of a fibration, where the fibres are 4d $K3$ surfaces
with orbifold singularities which vary over a 3-sphere. One then
obtains $G_2$-manifolds with orbifold singularities along the
sphere. For example, if the generic fibre has both an $SU(4)$ and an $SU(5)$ singularity,
then the $G_2$ manifold will have two
such singularities, both parametrized by disjoint copies of the sphere. In this case $N^1_i$
and $N^2_i$ are equal because $Q_1$ and $Q_2$ are in the same homology class,
which is precisely the special case that we consider above.

A similar picture arises from the dual perspective of the heterotic string on
a $T3$-fibred Calabi-Yau.
Then, if the hidden sector $E_8$ is broken by the background gauge field to, say, $SU(5) \times SU(2)$
the $K3$-fibers of the dual $G_2$-manifold generically have $SU(5)$ and $SU(2)$ singularities, again with
$N^1_i$ = $N^2_i$ (or $N^1_i = k N^2_i$ in general).

Finally, we note that Joyce's examples typically can have several sets of orbifold singularities which
often fall into the special class we have considered.

\noindent{\bf 4. Phenomenology}\\
As mentioned in section 2.2, there are many local minima with spontaneously broken susy.
One can study the particle physics features of these minima. For illustration, we will compute some phenomenologically relevant quantities
for the minima (\ref{sol7}):
\begin{eqnarray} \label{phenoquant}
&&V_0^{(3),(4)}\approx-(5.1\times10^{10}\,\mathrm{GeV})^4\;\footnotesize{\rm{(cosmological\;constant)}} \nonumber \\
&&m_{3/2}^{(3),(4)}=m_pe^{K/2}|W|\approx2081\,\mathrm{GeV\;\footnotesize{(gravitino\;mass)}} \nonumber \\
&&M_{11}=\frac{\sqrt{\pi}m_p}{V_X^{1/2}}\approx3.9\times 10^{17}\, \mathrm{GeV\;\footnotesize{(11\,dim\;Planck\;scale)}} \nonumber \\
&&\Lambda^{(1)}_{g}= m_p\,e^{-\frac{b_1}{3}\Sigma_iN_is^i}\approx2.6\times 10^{15}\, \mathrm{GeV}\nonumber\\
&&\Lambda^{(2)}_{g}\approx9.7\times10^{14}\,\mathrm{GeV\;\;\;\;\footnotesize{(gaugino\; cond.\; scales)}} \, ,
\end{eqnarray}
\noindent where $m_p=2.43\times 10^{18}$GeV and the hidden sector
strong coupling scales are defined as in
\cite{Intriligator:1995au}. Thus, standard gauge
unification is naturally compatible with low scale SUSY in our
theory. An investigation of the entire `parameter' set shows that
a significant fraction of models have similar features
\cite{Toappear}. Note that to obtain much lower mass scales
requires unnaturally large rank gauge groups \emph{and} large
$A_2/A_1$ ratios. Presumably, the latter cannot reach, say,
$\mathcal{O}(100)$, implying \emph{a lower bound on the susy
breaking scale} in these vacua.

A large negative $V_0$ is not realistic and one
might worry that the features obtained above may not survive when
one obtains (or tunes) $V_0$ to the correct value. We argue that
this is not the case. First, there may exist mechanisms in
these vacua similar to the Bousso-Polchinski mechanism
\cite{Bousso:2000xa} in IIB. The $M$ theory dual of IIB fluxes in principle
`scan $V_0$' leaving a minimum very close to the one discussed here.

Furthermore, in \cite{Toappear} we have studied the vacuum structure
with additional non-perturbative contributions and hidden sector matter, as in \cite{Lebedev:2006qq}.
These can give rise to vacua with a completely different $V_0$, eg. dS
vacua, but with essentially identical phenomenology. Moreover, if one
assumes that the space of $G_2$ manifolds is such that one can finely scan
the constants ${A_i}$, then we have checked that it is possible to scan $V_0$ to small values without
changing the phenomenology.

\noindent{\bf 4.1 Soft supersymmetry breaking parameters}
\\
Soft supersymmetry breaking parameters (at $M_{unif}$) can be
calculated in this framework - the gaugino masses $M^a_{1/2}$ are
easier to calculate than the scalars and trilinears. The MSSM
gaugino masses are given by: \be\label{ga1}
M_{1/2}=m_p\frac{e^{K/2}K^{i\bar{j}}F_{\bar{j}}\partial_{i}f_{sm}}{2{i\,\rm
Im}f_{sm}},\; f_{sm}=\sum_{i=1}^{N}N^{sm}_i z_i.\ee where the
$f_{sm}$ is determined by the homology class of the (MS)SM
3-cycle, $Q_{sm}$. From (\ref{ga1}), the normalized gaugino mass
in these vacua can be expressed as \be
\label{eq9}|M_{1/2}|=\left[\frac{2(\alpha
b_1-b_2)}{3(\alpha-1)}\frac{\sum_{i=1}^{N}{N^{sm}_is_i\nu_i}}
{\sum_{i=1}^{N}{N^{sm}_is_i}}+1\right]\times m_{3/2} \ee At the
SUSY extremum, using equation (\ref{eq8}) for $\nu_i$ in
(\ref{eq9}), $M_{1/2}$ vanishes as expected implying a perfect
cancellation between the two terms. The moduli for the AdS minima
with spontaneously broken susy are such that there is a subtle
cancellation (albeit not perfect) between the two terms, leading
to a suppression of the gauginos relative to $m_{3/2}$. This will
be explained further in \cite{Toappear}.

For the illustrative two moduli case with pure SYM hidden sector,
take $N^{sm}_1=2\,,\,\,\,\,N^{sm}_2=1$, so that the (MS)SM gauge
kinetic function is $f_{sm}=2 z_1+ z_2.$ The gaugino masses
 for the two vacua in (\ref{sol7}) are then \be \label{gaugino} |M_{1/2}^{(3),(4)}| =
m_p |\frac{e^{K/2}K^{i\bar{j}} F_{\bar{j}}
\partial_i f_{sm}}{2\,{\rm Im}f_{sm}}|\approx\{165,\,97\}\,\mathrm{GeV}.\ee Similar values arise for a significant
fraction of the parameters.
The tree level gaugino masses are universal but
the non-universal one-loop
anomaly mediated contributions are also non-negligible.

With $V_0$ tuned, the scalar masses are equal to $m_{3/2}$ times a
factor which is generically unsuppressed in these vacua, so
the scalar masses are expected to be of
$O(m_{3/2})$ - heavier than the gauginos. The trilinears (with the
yukawas factored out) turn out to be $ \geq m_{3/2}$. Since the
scalars are heavier than the gauginos, the LSP is a neutralino.

(\ref{gaugino}) gives a renormalized gluino mass of about \{500,
300\} GeV at the TeV scale and will give a clear signal at the LHC
beyond the standard model background. Eg, there will be an excess
of events with two charged leptons, at least two jets with a
transverse momentum greater than 100GeV and a large missing energy
from the LSP. This signal will be seen even with low luminosity.

The fact that the gaugino masses are suppressed, but the scalars
are not implies that LHC data could distinguish these vacua from
the KKLT models of Type IIB \cite{Choi:2005ge}. Some large volume
Type IIB vacua may give a spectrum similar to $M$ theory
\cite{Conlon:2006us}, but we expect that a more thorough study
\cite{Toappear} eg of the trilinears, will show that LHC is
capable of distinguishing these also.

{\noindent \bf 5. Remarks and Conclusions}\\
The stabilization of moduli and the hierarchy by strong dynamics in $M$ theory seems to be quite generic and robust.
The electroweak scale emerges from the fundamental theory even though the fundamental scale
and compactification scale are much larger.
Focussing on mechanisms which stabilize the hierarchy was useful
and complimentary to the approach of `searching for the Calabi-Yau which gives the MSSM spectrum
at the GUT scale'.
The $\mu$
problem, electroweak symmetry breaking, flavor and CP physics, dark matter, inflation and LHC physics can
all be addressed within this framework and
some of these studies are underway \cite{Toappear}. \\
{\noindent \bf Acknowledgements}
We would like to thank F. Larsen, T. Taylor and J. Wells for discussions. The work of GK, PK and DV is
supported in part by the DOE.
\bibliography{apssamp}

\begin{thebibliography}{15}
\bibitem{Giddings:2001yu}
S.~B.~Giddings, S.~Kachru and J.~Polchinski,
Phys.\ Rev.\ D {\bf 66}, 106006 (2002)
[arXiv:hep-th/0105097].
\bibitem{Kachru:2003aw}
S.~Kachru, R.~Kallosh, A.~Linde and S.~P.~Trivedi,
Phys.\ Rev.\ D {\bf 68}, 046005 (2003)
[arXiv:hep-th/0301240].
\bibitem{Acharya:2002kv}
  B.~S.~Acharya,
  arXiv:hep-th/0212294.
\bibitem{DeWolfe:2005uu}
  O.~DeWolfe, A.~Giryavets, S.~Kachru and W.~Taylor,
  JHEP {\bf 0507}, 066 (2005)
  [arXiv:hep-th/0505160].
  J.~P.~Derendinger, C.~Kounnas, P.~M.~Petropoulos and F.~Zwirner,
  Nucl.\ Phys.\ B {\bf 715}, 211 (2005)
  [arXiv:hep-th/0411276].
\bibitem{Gukov:2003cy}
  S.~Gukov, S.~Kachru, X.~Liu and L.~McAllister,
  Phys.\ Rev.\ D {\bf 69}, 086008 (2004)
  [arXiv:hep-th/0310159].


\bibitem{Randall:1999ee}
  L.~Randall and R.~Sundrum,
  Phys.\ Rev.\ Lett.\  {\bf 83}, 3370 (1999)
  [arXiv:hep-ph/9905221].

\bibitem{Acharya:1998pm}
  B.~S.~Acharya,
  Adv.\ Theor.\ Math.\ Phys.\  {\bf 3} (1999) 227
  [arXiv:hep-th/9812205]
;  B.~S.~Acharya,
  arXiv:hep-th/0011089.


\bibitem{Acharya:2001gy}
  B.S.~Acharya and E.~Witten,
  arXiv:hep-th/0109152.

\bibitem{Acharya:2004qe}
  B.~S.~Acharya and S.~Gukov,
  Phys.\ Rept.\  {\bf 392} (2004) 121
  [arXiv:hep-th/0409191].

\bibitem{Atiyah:2001qf}
  M.~Atiyah and E.~Witten,
  Adv.\ Theor.\ Math.\ Phys.\  {\bf 6}, 1 (2003)
  [arXiv:hep-th/0107177].

\bibitem{Witten:2001bf}
  E.~Witten,
  arXiv:hep-ph/0201018.

\bibitem{Friedmann:2002ty}
  T.~Friedmann and E.~Witten,
  Adv.\ Theor.\ Math.\ Phys.\  {\bf 7}, 577 (2003)
  [arXiv:hep-th/0211269].

\bibitem{Acharya:2005ks}
  B.~S.~Acharya and R.~Valandro,
  arXiv:hep-ph/0512144.

\bibitem{Harvey:1999as}
  J.~A.~Harvey and G.~W.~Moore,
  arXiv:hep-th/9907026.

\bibitem{Acharya:2005ez}
  B.~S.~Acharya, F.~Denef and R.~Valandro,
  JHEP {\bf 0506}, 056 (2005)
  [arXiv:hep-th/0502060].

\bibitem{Toappear} B. Acharya, K. Bobkov, G. Kane, P. Kumar, D. Vaman, {\it To appear.}

\bibitem{deCarlos:1992da}
  B.~de Carlos, J.~A.~Casas and C.~Munoz,
  Nucl.\ Phys.\ B {\bf 399}, 623 (1993)
  [arXiv:hep-th/9204012].
  J.~J.~Blanco-Pillado, R.~Kallosh and A.~Linde,
  JHEP {\bf 0605}, 053 (2006)
  [arXiv:hep-th/0511042].
  H.~Abe, T.~Higaki and T.~Kobayashi,
  Nucl.\ Phys.\ B {\bf 742}, 187 (2006)
  [arXiv:hep-th/0512232].
  D.~Krefl and D.~Lust,
  JHEP {\bf 0606}, 023 (2006)
  [arXiv:hep-th/0603166].


\bibitem{ddf}
  F.~Denef, M.~R.~Douglas and B.~Florea,
  JHEP {\bf 0406}, 034 (2004)
  [arXiv:hep-th/0404257].

\bibitem{web}
{\it http://feynman.physics.lsa.umich.edu/$\sim$signaturespace.}

\bibitem{Kovalev:2001zr}
 D.D. Joyce, `Compact Manifolds with Special Holonomy,'' Oxford University Press, 2000;
 A.~Kovalev,
  arXiv:math.dg/0012189.

\bibitem{Intriligator:1995au}
  K.~A.~Intriligator and N.~Seiberg,
  Nucl.\ Phys.\ Proc.\ Suppl.\  {\bf 45BC}, 1 (1996)
  [arXiv:hep-th/9509066].

\bibitem{Bousso:2000xa}
  R.~Bousso and J.~Polchinski,
  JHEP {\bf 0006}, 006 (2000)
  [arXiv:hep-th/0004134].

\bibitem{Lebedev:2006qq}
  O.~Lebedev, H.~P.~Nilles and M.~Ratz,
  Phys.\ Lett.\ B {\bf 636}, 126 (2006)
  [arXiv:hep-th/0603047].

\bibitem{Conlon:2006us}
  J.~P.~Conlon and F.~Quevedo,
  arXiv:hep-th/0605141.
\bibitem{Choi:2005ge}
  K.~Choi, A.~Falkowski, H.~P.~Nilles and M.~Olechowski,
  Nucl.\ Phys.\ B {\bf 718}, 113 (2005)
  [arXiv:hep-th/0503216].
  R.~Kitano and Y.~Nomura,
  Phys.\ Lett.\ B {\bf 631}, 58 (2005)
  [arXiv:hep-ph/0509039].
  A.~Pierce and J.~Thaler,
  arXiv:hep-ph/0604192.
  H.~Baer, E.~K.~Park, X.~Tata and T.~T.~Wang,
  arXiv:hep-ph/0604253.
\end{thebibliography}

\end{document}